\documentclass[creativecommons]{eptcs}

\providecommand{\event}{ACL2-2022}

\usepackage{alltt}
\usepackage{enumitem}
\usepackage{listings}
\usepackage{underscore}
\usepackage{url}
\usepackage{xcolor}

\newcommand{\acl}[1]{\texttt{#1}} 
\newenvironment{bacl}{\small\begin{alltt}}{\end{alltt}} 

\newcommand{\syn}[1]{\texttt{#1}} 

\newcommand{\secref}[1]{Section \ref{#1}}

\newcommand{\citecode}[2]{\cite[Path \href{#1}{\texttt{#2}}]{acl2-code}}
\newcommand{\citeman}[2]{\cite[Topic \href{#1}{\texttt{#2}}]{acl2-manual}}

\newcommand{\todo}[1]{}

\lstdefinelanguage{syntheto}
  {morekeywords={function,return,returns,forall,struct,variant,theorem,specification,if,else,let,assumes,ensures,transform,by,subtype},
   sensitive=true,
   morecomment=[l]{//},
   morecomment=[s]{/*}{*/},
   morestring=[b]",
  }

\begin{document}


\title{Syntheto: A Surface Language for APT and ACL2%
       \thanks{US Government Notice:
               Approved for Public Release, Distribution Unlimited}}

\author{Alessandro Coglio
        \quad\quad
        Eric McCarthy
        \quad\quad
        Stephen Westfold
        \institute{Kestrel Institute \\
                   \url{http://www.kestrel.edu}}
        \and
        Daniel Balasubramanian
        \quad\quad
        Abhishek Dubey
        \quad\quad
        Gabor Karsai
        \institute{Institute for Software-Integrated Systems,
                   Vanderbilt University \\
                   \url{https://www.isis.vanderbilt.edu}}}

\def\titlerunning{Syntheto}
\def\authorrunning{A. Coglio, E. McCarthy, S. Westfold,
                   D. Balasubramanian, A. Dubey, G. Karsai}

\maketitle

\begin{abstract}
Syntheto is a surface language for carrying out
formally verified program synthesis by transformational refinement
in ACL2 using the APT toolkit.
Syntheto aims at providing more familiarity and automation,
in order to make this technology more widely usable.
Syntheto is a strongly statically typed functional language
that includes both executable and non-executable constructs,
including facilities to state and prove theorems
and facilities to apply proof-generating transformations.
Syntheto is integrated into an IDE with a notebook-style, interactive interface
that translates Syntheto to ACL2 definitions and APT transformation
invocations, and back-translates the prover's results to Syntheto;
the bidirectional translation happens behind the scenes,
with the user interacting solely with Syntheto.
\end{abstract}



\section{Introduction}
\label{sec:intro}

The APT (Automated Program Transformations) toolkit \cite{apt-www},
built on the ACL2 theorem prover \cite{acl2-www},
provides facilities to carry out
\emph{formally verified program synthesis by transformational refinement}.
Using APT involves using ACL2,
and requires expertise in both ACL2 and APT.

In the pursuit of the worthy but arduous goal
to make APT and ACL2 (and more generally, formal methods) more widely usable,
we have designed and developed a prototype of Syntheto,
a \emph{surface language} for ACL2 and APT.
Syntheto contains constructs for writing formal specifications
and for applying proof-generating transformations to them.
These constructs are translated to ACL2 definitions and APT transformation
invocations, and the prover's results are translated back to Syntheto.

Compared to ACL2 and APT,
Syntheto aims at providing more \emph{familiarity} and \emph{automation}.
Its \emph{syntax} is more similar to popular programming languages like Java,
with curly braces and infix operators.
Syntheto is \emph{strongly statically typed}:
this feature is commonly found in popular programming languages,
and helps automate a decidable portion of
what manifests as potentially undecidable guard obligations in ACL2.
Moreover, the typing facilitates termination proofs,
by effectively removing from consideration
function arguments outside their types (i.e.\ guards in ACL2).%
\footnote{We are not taking a position on
whether Lisp-like syntax is better or worse than Java-like syntax,
or whether untyped languages are better or worse than typed languages;
in fact, our position is simply that there are relative pros and cons.
We are merely stating that, based on our observations,
the majority of users, both inside and outside the formal methods community,
finds certain language features more familiar and intuitive than others.
This motivates our design choices for Syntheto.}
Syntheto transformations also increase automation
by \emph{combining multiple APT transformations} in some cases;
however, similar combinations could be realized in ACL2 and APT,
and are therefore not necessarily a unique characteristic of Syntheto.

Last but not least, Syntheto is integrated into an IDE
with a notebook-style \cite{PER-GRA:2007},
interactive interface,
which in particular continuously displays the results of transformations
in read-only portions just below the invocations of the transformations,
which are refreshed (i.e.\ re-calculated)
when the user edits upstream code.
This is particularly helpful in multi-step program synthesis,
because the user can readily see
the intermediate specifications produced by a sequence of transformations
just below the transformations that produce them
and can quickly assess the impact of alternative choices.

Syntheto is no silver bullet:
it does not make APT and ACL2 suddenly usable
by anyone without a formal methods background.
Nonetheless, our initial experience suggests that
it may be easier to approach for more users than plain APT and ACL2.
Syntheto is currently just a prototype,
as we had limited time and resources to develop it;
more development is needed to draw
more solid conclusions about the merits of Syntheto.
In terms of contributions, we also believe that
the bidirectional translation between Syntheto and ACL2/APT
involves interesting technical aspects.

\secref{sec:language} overviews the Syntheto language.
\secref{sec:frontend} describes the \emph{Syntheto front end},
i.e.\ the notebook-style IDE
with which the user interacts.
\secref{sec:backend} describes the \emph{Syntheto back end},
i.e.\ the machinery that translates between Syntheto and ACL2/APT,
as well as all the supporting developments.
\secref{sec:examples} provides an example of use of Syntheto.
\secref{sec:related} discusses related work.
\secref{sec:future} outlines future work
to further develop Syntheto and overcome its current limitations.

Syntheto is available in the ACL2 Community Books,
at \acl{[books]/kestrel/syntheto}.


\section{Language}
\label{sec:language}

Syntheto is a strongly statically typed purely functional language
that includes non-executable constructs for specification,
as well as constructs for stating theorems
and constructs for transforming functions.

\subsection{Types}
\label{sec:language-types}

Syntheto has built-in primitive types for
booleans (the usual two),
characters (ISO 8851-1, 8-bit),
strings (of such characters, of any length),
and integers (unbounded).
It also has built-in parameterized types for options,%
\footnote{An option type is a disjoint union of a base type
with a distinct value that indicates the absence of a value of the base type.
An option type can represent an optionally present value of the base type.}
finite sets,
finite sequences,
and finite maps:
the first one is parameterized over the base type,
the second and third ones are parameterized over the element type,
and the fourth one is parameterized over the domain and range types.

Syntheto supports user-defined named algebraic types.
A \emph{product type} is defined by a set of named type components
and an optional invariant boolean expression over the component names:
a value of a product type is a finite map
from the component names to values of their types
that satisfy the invariant.
A \emph{sum type} is defined by a set of named alternatives
each of which has an associated unnamed product type:
a value of a sum type consists of
the name of an alternative and a value of the corresponding product type. An example of a
product type and a sum type:
\begin{lstlisting}
struct rational {
   numerator: int,
   denominator: positive
  | gcd(abs(numerator), abs(denominator)) == 1
}
variant orientation {clockwise, counterclockwise, colinear}
\end{lstlisting}

Syntheto also supports user-defined named subset types (i.e.\ subtypes).
A \emph{subtype} is defined by an (existing) supertype and
a restricting boolean expression over the supertype:
the subtype consists of
the values of the supertype that satisfy the restriction.
Since Syntheto types must be non-empty,
the definition of a subtype also includes an optional ground expression
that must evaluate to a witness value of the subtype,
i.e.\ a value of the supertype that satisfies the restriction;
this expression must be supplied by the user
when a witness value cannot be automatically inferred.

Recursive types, including mutually recursive ones, are supported.
The declaration of a clique of mutually recursive types
must be enclosed in a surrounding declaration
that explicitly groups the clique.
Type recursions must be well-founded;
this requirement is handled via theorem proving.

\subsection{Expressions}
\label{sec:language-expressions}

Syntheto has
literals
(of the four primitive types mentioned in \secref{sec:language-types}),
typed variables,
unary and binary expressions for a variety of operators,
ternary conditional expressions,
function calls,
local variable bindings,
tuple constructors and accessors,
as well as constructors, accessors, and updaters
of values of product and sum types.
Being a purely functional language, Syntheto has no statements;
however, the syntax of certain kinds of expressions
resembles statement syntax more than expression syntax.

Boolean expressions are also used as logical formulas.
Boolean operators include not only negation, conjunction, and disjunction,
but also implication, converse implication, and coimplication.

Establishing the type correctness of Syntheto expressions
requires type checking and some amount of type inference
(the latter is for the built-in parameterized types
mentioned in \secref{sec:language-types}).
Type checking involves proof obligations,
e.g.\ that a product type construction expression satisfies the invariant,
and that an expression of a supertype
to which a variable of a subtype is bound
satisfies the subtype restriction.
Syntheto type checking is thus split into
a decidable portion and an undecidable portion,
similarly to languages like
PVS \cite{pvs-www} and Specware \cite{specware-www}:
the decidable portion is handled by an algorithm,
while the undecidable portion engenders proof obligations
that are handled via theorem proving
(as explained in more detail later).

\subsection{Functions}
\label{sec:language-functions}

Syntheto has two kinds of function definitions,
both of which start with a function header
that consists of a name,
zero or more named typed inputs,
and one or more named typed outputs.
There may also be a precondition,
i.e.\ a boolean expression over the inputs,
and a postcondition,
i.e.\ a boolean expression over the inputs and the outputs.
The precondition engenders a proof obligation
for each call of the function.
The postcondition and precondition engender a proof obligation
on the function definition itself.

The first kind of function definition (\emph{regular})
has a body consisting of an expression over the inputs,
which calculates the outputs from the inputs.
This kind of function is executable
if all the functions it calls are executable.
This kind of function may be singly or mutually recursive;
mutually recursive cliques of functions
must be enclosed in a surrounding declaration
that explicitly groups the clique. An example regular definition of the factorial function:
\begin{lstlisting}
function factorial (n:int) assumes n >= 0 returns (out:int) ensures out > 0 {
  if (n == 0) {
    return 1;
  }
  else {
    return n * factorial(n - 1);
  }
}
\end{lstlisting}

The second kind of function definition (\emph{quantified}) consists of a universal or
existential quantification, over one or more named typed variables (distinct from inputs
and outputs), and a boolean expression over input and quantified variables.  This kind of
function definition must have one output of boolean type.  Functions with quantifiers
cannot be recursive.

Syntheto also has some built-in functions
over sets, sequences, and other built-in types.
These do not have an explicit definition
of the two kinds discussed above;
they have implicit definitions.
There is currently a limited selection of built-in functions;
more built-in functions may be added with ease.

\subsection{Specifications}
\label{sec:language-specifications}

A Syntheto function of the kinds described above
can be used as a specification of an input/output behavior.
The function may be non-executable, or inefficiently executable;
its role is just to specify an input/output behavior in a simple and clear way.
This specification may be refined, via Syntheto transformations,
to another function that is efficiently executable,
and likely more complicated than the initial specification:
the new function is the implementation of the specification.
The specification and implementation functions
are extensionally equal but intensionally different:
they denote the same mathematical function, but have different bodies.

Syntheto also supports more general specifications,
in the form of second-order predicates.
A Syntheto second-order predicate is over one or more function variables,
which are placeholders for the efficiently executable functions
to be synthesized via transformational refinement.
The predicate states constraints that must be satisfied by the functions,
in terms of their inputs and outputs (i.e.\ extensionally),
not in terms of their bodies (i.e.\ intensionally).
This kind of Syntheto specification may be refined to a form in which
the function variables are equated to efficiently executable Syntheto functions
that are synthesized in the course of the transformational refinement.
These Syntheto functions satisfy the predicate by construction;
they are solutions to the constraint problem expressed by the predicate.

The simpler kind of Syntheto specification, discussed above,
that consists of a single function
(or more generally, a clique of mutually recursive functions)
can be viewed as an abbreviation for a second-order predicate
that constrains a function variable to be
(extensionally) equal to the Syntheto function.

A Syntheto second-order predicate has a name,
and takes one or more function headers as arguments.
A function header, as mentioned in \secref{sec:language-functions},
consists of a function name and of named typed inputs and outputs:
when used as argument of a second-order predicate,
the function name is the name of the function variable,
and the named inputs and outputs can be used
to express input/output constraints.
The body of a second-order predicate may be one of three possible kinds.
The first kind is a boolean expression over the function variables,
which may reference other predicates.
The second kind consists of
a universal or existential quantification
over one or more named typed variables,
and a boolean expression over
the function variables and the quantified variables,
which may reference other predicates.
Note that these two kinds are analogous to
the two kinds of function definitions
discussed in \secref{sec:language-functions}.
The third kind, only usable for predicates over single function variables,
is a boolean expression over
the named inputs and outputs of the function header:
this covers the common case of an input/output relation
that a synthesized function must specify.
This third kind can be viewed as a special case of the second kind,
where the function inputs and outputs are universally quantified.
The names of the inputs and outputs in the function headers
are only used in this third kind of specification;
in the other two kinds,
only the names of the function variables are used.
An example specification (of the third kind) of a sort function:
\begin{lstlisting}
specification sort_spec
  (function sort (input: seq<int>) returns (output: seq<int>)) {
   ordered(output) && permutation(output, input);
}
\end{lstlisting}

\subsection{Theorems}
\label{sec:language-theorems}

Syntheto has a construct to declare, and attempt to prove, explicit theorems.
The explicit theorems are in addition to implicit theorems that arise from
the type checking proof obligations described in \secref{sec:language-expressions}
and the function precondition/postcondition proof obligations
described in \secref{sec:language-functions}.

An explicit Syntheto theorem consists of a name,
a sequence of typed variables,
and a boolean expression (i.e.\ the formula) over those variables.
The variables are universally quantified.
We plan to extend this construct with the ability to specify proof hints.

\subsection{Transformations}
\label{sec:language-transformations}

Syntheto transformations generate a definition of a function in terms of an existing
function. The transformation to apply has a name with named options specified using product
syntax. 
The following transformations are currently supported:
\begin{itemize}[nosep]
 \item{\syn{simplify}}: Simplifies a function definition using enabled rewrite rules.
 \item{\syn{finite_difference}}: Adds a parameter to a function along with an invariant
   that the parameter is equal to a function of the existing
   parameters, to incrementally compute an expensive expression.
  \item{\syn{tail_recursion}}: Puts a function into tail-recursive form.
  \item{\syn{rename_param}}: Renames a parameter.
  \item{\syn{isomorphism}}: Replaces a parameter of one type by a parameter of an
    isomorphic type.
  \item{\syn{drop_irrelevant_parameter}}: Removes a parameter that is not needed in
    computing the result of the function---usually because of a previous finite-difference
    transformation.
  \item{\syn{wrap_output}}: Wraps a function call around the body of a function.
  \item{\syn{restrict}}: Adds a precondition on a function. This may enable another
    transformation such as \syn{isomorphism}.
\end{itemize}

\subsection{Executability}
\label{sec:language-executability}

Syntheto has both executable and non-executable constructs,
as appropriate for a formal specification language.
The Syntheto expressions described in \secref{sec:language-expressions}
are all executable, provided that every function they call is executable.
The functions described in \secref{sec:language-functions}
whose definitions are expressions are executable,
provided that every function called by those expressions are executable;
the functions with universal or existential quantifiers are not executable.
The specification predicates described in \secref{sec:language-specifications}
are not executable.
Explicit and implicit theorems are not executable.
Transformations are executable;
they may generate executable or non-executable functions.


\section{Front End}
\label{sec:frontend}

With an interactive theorem prover, admissibility of definitions and
provability of theorems are sensitive to the current theory. It is common to
build proofs in a linear manner. To make this familiar to the majority of
users, we use a \emph{notebook-style interface} \cite{PER-GRA:2007}.

When starting with an empty notebook, definitions are appended one by one, and
the front end submits them, in order, to the ACL2 back end.  A new definition
cannot be added if the previous submission was rejected---rather, the rejected
definition must first be fixed and resubmitted.  If an earlier cell is edited,
then it and all definitions after it are resubmitted in order, until there is
either a rejected definition or until all cells have been submitted and accepted.

The front end components are described here because they are conceptually
important for understanding and usability. We describe how these
components work in our prototype.

\subsection{Components}
\label{sec:components}

From a user's point of view, the main components needed to use Syntheto are
\begin{itemize}[nosep]
\item Visual Studio Code (VS Code) \cite{vscode},
\item Syntheto plugin for VS Code,
\item Docker Engine \cite{docker-engine}, and
\item a Docker image containing the Syntheto back end.
\end{itemize}

\noindent
We briefly describe the subcomponents of the Syntheto plugin for VS Code
and how they are built.  In later sections we describe in more detail what
happens when a user enters a definition using the plugin.

The Syntheto parser is generated from an Xtext grammar \cite{xtext}.  The
generator produces Java code that parses the Syntheto code and builds Xtext
abstract syntax trees (ASTs).  Another Java package converts those ASTs to a
form that is homomorphic to the Syntheto abstract syntax definition in ACL2,
which we call the Syntheto ASTs.  The VS Code Notebook interface and Language
Server Protocol (LSP) interface are also written in Java.

The \texttt{edu.kestrel.syntheto.ast} Java package defines Syntheto ASTs.
In general, each class represents a language construct.
Besides defining the fields of the construct, each class defines methods to
convert instances to and from S-expression AST structures.

The \texttt{edu.kestrel.syntheto.sexpr} Java package defines ACL2
S-expression ASTs. This model of S-expressions is designed to be easily
serialized to text that is then easily parsed by ACL2. For translating code
in the other direction, this package also defines token classes, a tokenizer
that lexes a stream of characters into a stream of tokens, and a parser that
constructs an S-expression AST from a sequence of tokens.

\subsection{Setup}
\label{sec:setup}

A user starts the Docker image that contains the customized ACL2 executable,
which starts listening on an IP socket or Unix domain socket. The user then
starts VS Code with the Syntheto plugin. The VS plugin recognizes the
\texttt{.synth} file extension. When the user opens a new or existing file
with that extension, the plugin establishes a connection with the socket.

\subsection{Parsing}
\label{sec:parsing}

When the user enters a definition or theorem into a cell and presses the
\textbf{run} button, the text is parsed and restructured by the front end
into a Syntheto AST in Java. The AST classes correspond to the sum and product
fixtypes that define the Syntheto ASTs in ACL2.

For a Syntheto construct defined by an ACL2 product fixtype
\cite{15-swords-fty}
\citeman{https://www.cs.utexas.edu/users/moore/acl2/manuals/latest/index.html?topic=FTY____DEFPROD}{defprod}
we define a concrete subclass
with fields for the formal parameters of the language construct.
For a Syntheto construct defined by an ACL2 sum fixtype
\cite{15-swords-fty}
\citeman{https://www.cs.utexas.edu/users/moore/acl2/manuals/latest/index.html?topic=FTY____DEFTAGSUM}{deftagsum}
we define an abstract superclass;
for each alternative component of the sum type, we define a
concrete subclass with fields for the formal parameters of the component.

\subsection{Transfer Language}
\label{sec:transfer-language}

In order to communicate between the Java front end and the ACL2 back end, we
define a simple S-expression language called \emph{transfer language}.
The transfer language is chosen to be a subset of ACL2 syntax that is simple
for Java to parse. For example, it does not support Lisp reader macros or backquote.

The transfer language is interpreted in the \texttt{ACL2} package, meaning
symbols with no package prefix refer to symbols accessible in the \texttt{ACL2}
package. The syntax for symbols is very simple: only uppercase letters,
periods, colons, and hyphens are allowed. Vertical bar and backslash escapes
are not allowed in the language (although for expediency, for now
they do not trigger runtime errors when encountered).

For example, consider the Syntheto source code:
\begin{lstlisting}
subtype positive {
  x: int | x > 0
}
\end{lstlisting}

The transfer language form sent to the back end is:
\begin{bacl}
(SYNTHETO::PROCESS-SYNTHETO-TOPLEVEL
 (SYNTHETO::MAKE-TOPLEVEL-TYPE
  :GET (SYNTHETO::MAKE-TYPE-DEFINITION
        :NAME (SYNTHETO::MAKE-IDENTIFIER :NAME "positive")
        :BODY (SYNTHETO::MAKE-TYPE-DEFINER-SUBSET
               :GET (SYNTHETO::MAKE-TYPE-SUBSET
                     :SUPERTYPE (SYNTHETO::MAKE-TYPE-INTEGER)
                     :VARIABLE (SYNTHETO::MAKE-IDENTIFIER :NAME "x")
                     :RESTRICTION
                     (SYNTHETO::MAKE-EXPRESSION-BINARY
                      :OPERATOR (SYNTHETO::MAKE-BINARY-OP-GT)
                      :LEFT-OPERAND
                      (SYNTHETO::MAKE-EXPRESSION-VARIABLE
                       :NAME (SYNTHETO::MAKE-IDENTIFIER :NAME "x"))
                      :RIGHT-OPERAND
                      (SYNTHETO::MAKE-EXPRESSION-LITERAL
                       :GET (SYNTHETO::MAKE-LITERAL-INTEGER :VALUE 0)))
                     :WITNESS NIL)))))
\end{bacl}

The transfer language form returned by the back end is:
\begin{bacl}
(SYNTHETO::MAKE-OUTCOME-TYPE-SUCCESS :MESSAGE "positive")
\end{bacl}

In most cases, a form in the transfer language consists of an outermost macro
wrapped around a tree of ACL2 fixtype \texttt{MAKE-} forms. We call the inner
forms \emph{make-myself} forms. Every ACL2 sum or product fixtype used for
Syntheto ASTs has an associated \texttt{<typename>--MAKE-MYSELF} function that
takes an AST instance argument and returns ACL2 code that, when
executed, makes an identical AST instance.

Syntheto definition names and other names in Syntheto syntax have a different
syntax from ACL2 symbols. For example, they are case sensitive.
To avoid using symbols with escaped lowercase letters, the definitions are represented
in the AST as strings, not symbols. This can be seen in forms such as:
\begin{bacl}
(SYNTHETO::MAKE-IDENTIFIER :NAME "x").
\end{bacl}

In the back end, Syntheto definition names are eventually translated to ACL2 definition
names, which are symbols. To avoid problems that would arise when a name
already has a definition in a different package, we define the \texttt{SYNDEF}
package, which does not import any symbols from any other package. Syntheto
definition names, field names, composite type names, parameter names, and
local variable names are all interned in the \texttt{SYNDEF} package. Derived
symbols such as \texttt{sequence[int]-p} and names of automatically generated
theorems are interned in the package of the symbol they were derived from, so
they are also in the \texttt{SYNDEF} package.

\subsection{Java AST to S-Expression AST}
\label{sec:ast-conversion}

Following parsing, the Syntheto AST is transformed to an S-expression AST that
represents a make-myself form, as described above. The S-expression AST
classes model ACL2 lists, characters, integers, strings, and symbols.

Since Java integers are limited in size, the S-expression class modeling ACL2
integers uses the Java \texttt{BigInteger} class.

The S-expression class for symbols has fields for the package name and the
symbol name. There are four \emph{special symbols} with package name
\texttt{COMMON-LISP}. They are: \texttt{T}, \texttt{NIL}, \texttt{LIST},
and \texttt{CODE-CHAR}.

\subsection{Serialization}
\label{sec:serialization}

S-Expression ASTs are serialized to text as ACL2 code. Characters and symbols
have some special serialization rules.

A class instance representing a character is serialized as a form that makes
the character when evaluated. For example, the character that ACL2 would
print as \verb$#\!$ is serialized as \texttt{(CODE-CHAR 33)}. This simplifies
the transfer language.

A class instance representing a symbol is serialized in one of three ways.
If the symbol is one of the special symbols mentioned above, it is
serialized without a package prefix. If the package is \texttt{KEYWORD}, the
package name is omitted and the symbol name is preceded by a single colon. Any
other symbol is serialized with the package name, two colons, and the symbol
name.

\subsection{ACL2 Bridge}
\label{sec:acl2-bridge}

The last thing that happens in Java after the user presses the
\textbf{run} button for a cell is that the wrapped, serialized S-expression
is sent to the ACL2 server over the ACL2 Bridge
\cite{13-davis-embedding}
\citeman{https://www.cs.utexas.edu/users/moore/acl2/manuals/latest/index.html?topic=ACL2____BRIDGE}{bridge}.

A wrapper is added to the serialized text that instructs the ACL2 Bridge what to
do. Since the ACL2 form generally affects the world state, the outermost
wrapper, \texttt{(try-in-main-thread ...)}, tells the ACL2 Bridge that the
contents should be executed in the main Lisp thread. This is because ACL2
uses thread-local variables to hold some state information, and it also serves to
serialize the events. The next wrapper is \texttt{(nld ...)}, which is a
noninteractive version of ACL2's \texttt{LD} intended to be used
programmatically. In the case of an error, \texttt{(nld ...)} saves
error information rather than actually signalling an error.

The Java code that connects to the ACL2 server port is part of the LSP
component, and was adapted from the \texttt{edu.kestrel.syntheto.bridge} classes.
After connecting to the ACL2 server, it formats the wrapped, serialized ACL2 form
according to the protocol defined by the ACL2 Bridge and the stream is
sent over the wire. On the back end, \texttt{nld} evaluates the given form
as if it were entered to the ACL2 read-eval-print loop (REPL),
and the ACL2 Bridge returns the result returned by \texttt{nld}.

\subsection{Results Returned}
\label{sec:results-returned}

The \texttt{PROCESS-SYNTHETO-TOPLEVEL} macro ensures that
the returned outcome object and any Syntheto AST in it are formatted
as make-myself forms, so that they can be easily parsed in Java.

The steps above are then reversed, up to the Syntheto source code equivalent of
any Syntheto ASTs returned. The results are written into a non-editable
field following the cell that was run by the user.

One interesting kind of output is the result of a transformation.
A transformation produces a new definition that is in the ACL2 world state.
This definition is translated back to Syntheto source code and displayed
in the IDE.


\section{Back End}
\label{sec:backend}


The back end translates Syntheto constructs into ACL2, submits them to ACL2,
records and returns the results, and also translates transformed ACL2
functions back into Syntheto functions. The translation ensures that the
strong typing of Syntheto is always reflected in the corresponding predicates
in ACL2.

\subsection{Types}

The primitive types (boolean, string, character, integer) are translated to the
corresponding ACL2 types. The non-primitive types are translated to corresponding
fixtypes: product types to \acl{fty::defprod}, sum types to
\acl{fty::deftagsum}, finite sets to \acl{fty::defset}, finite maps to \acl{fty::defomap},
finite sequences to \acl{fty::deflist}, option types to a kind of \acl{fty::deftagsum}, and
subtypes to \acl{fty::defsubtype}.

Currently the polymorphic Syntheto types must always be instantiated. Each top-level Syntheto
construct is analyzed to see if any such types appear and the necessary instances
are created. The types may be nested, so the instantiations are created from the bottom up.

For every type there is a corresponding predicate name. The mapping from types to
predicate names is invertible, which is important for back translation from ACL2 to Syntheto.

\subsection{Expressions}

The translation of Syntheto expressions into ACL2 expressions is straightforward, making
use of the functions created by the type definition macros.

\subsection{Functions}

Syntheto functions with regular and quantified definitions are translated
into ACL2 functions with \acl{defun} and \acl{defun-sk}. The types of the
function parameters and the precondition are used to create guards for the functions. The
types of the function outputs and the postcondition are translated into theorems.

Some care is taken to help ACL2 verify the guards and prove the measure theorem (if the
function is recursive) needed for ACL2 to accept the function definition, and to prove the
output theorems. A test of the guard is incorporated into the body (wrapped in \acl{mbt}
\citeman{https://www.cs.utexas.edu/users/moore/acl2/manuals/latest/index.html?topic=ACL2____MBT}{mbt}).
The values returned are arbitrary in the case this guard expression is false, but values are
chosen that satisfy the output types to help ACL2 with its proofs.

Simple heuristics are used to generate an effective measure expression. This has been
sufficient for our current examples, but future work is likely to incorporate more sophisticated
techniques such as those used in the ACL2 Sedan.

\subsection{Specifications}

Function specifications are translated using \acl{defstub} to introduce the ACL2 name and
signature, and \acl{defun-sk} with a universal quantification to constrain it
\citeman{https://www.cs.utexas.edu/users/moore/acl2/manuals/latest/index.html?topic=ACL2____DEFSTUB}{defstub}
\citeman{https://www.cs.utexas.edu/users/moore/acl2/manuals/latest/index.html?topic=ACL2____DEFUN-SK}{defun-sk}.

\todo{Put simple example here or refer to later example?}

\subsection{Theorems}

Syntheto theorems are translated naturally to ACL2 theorems with the addition of
type predicate hypotheses for the typed variables. However, \acl{remove-hyps}
is used to remove any of these added hypotheses that are not actually needed to prove the
theorem \citeman{https://www.cs.utexas.edu/users/moore/acl2/manuals/latest/index.html?topic=ACL2____REMOVE-HYPS}{remove-hyps}.

\subsection{Transformations}

A single Syntheto transformation may map to one or more ACL2 APT transformations,
depending on the options given. For example, the \acl{tail_recursion} transformations is
preceded by a simplification step that applies rules that can put the function in the
correct form for the main transformation to occur. Several transformations, such as
\acl{finite_difference} and \acl{isomorphism}, simplify the result of the main transformation,
which is almost always what is wanted, but with an option to suppress the
simplification. The translation to APT is specified using schemas, so support for
additional APT transformations is easy to add to Syntheto.

\subsection{Back Translation of Transformed Functions}

Translation of ACL2 functions back into Syntheto functions requires inferring the types of
variables and removing the typing and guard predicates that occur in the function
body. Types of parameters can be extracted from a function's guard, and simple type
inference is done on the body. This type inference and the translation of expressions
relies on the strict invertible naming conventions we used for the Syntheto-to-ACL2
translation. 

During the translation of the body, all typing expressions are treated as \syn{true} and
standard simplifications are applied to remove them. The result is that a large
complicated-looking ACL2 definition can result in a significantly smaller, simpler
Syntheto definition.

We designed the translation of Syntheto to ACL2 so that it is invertible, but it is
possible that transformations could introduce functions that cannot be translated back to
Syntheto, or they could lose information necessary for inferring Syntheto types. For the
transformations we have currently linked to Syntheto, it is only necessary to prevent
simplification from unfolding certain definitions.


\section{Example: Point in Polygon Specification and Optimization}
\label{sec:examples}

The example we will use to illustrate Syntheto is the problem of finding whether a point
is inside a polygon. First the domain model is defined. This includes points, edges,
connectedness of edges, paths, whether edges intersect, and polygons. As well as
definitions, this model includes basic theorems about their properties. The example
algorithm specification considers an edge from the point of interest to a point known to
be outside the polygon: the point is in the polygon if and only if there are an odd number of crossings of this edge with the
edges representing the sides of the polygon. The full
specification and derivation is in
\citecode{https://github.com/acl2/acl2/blob/master/books/kestrel/syntheto/examples/point_in_polygon.synth}{[books]/kestrel/syntheto/examples/point_in_polygon.synth}.

Points and edges are defined with Syntheto product types:
\begin{lstlisting}
struct point {
  x: int,
  y: int
}
struct edge {
  p1: point,
  p2: point
}
\end{lstlisting}
These translate directly into product types in ACL2:
\begin{bacl}
(fty::defprod point
  ((x int) (y int))
  :tag :point)
(fty::defprod edge
  ((p1 point) (p2 point))
  :tag :edge)
\end{bacl}

Two edges are connected if the end point of the first is the start point of the
second. A path is a connected sequence of edges. The theorem states that the
tail of a non-empty path is also a path.
\begin{lstlisting}
function connected(e1:edge, e2:edge) returns (b:bool) {
  return e1.p2 == e2.p1;
}
function path_p(edges:seq<edge>) returns (b:bool) {
  return length(edges) <= 1
        || (connected(first(edges), first(rest(edges)))
             && path_p(rest(edges)));
}
theorem path_p_rest
  forall(edges:seq<edge>)
    !is_empty(edges) && path_p(edges)
      ==> path_p(rest(edges))
\end{lstlisting}

These Syntheto definitions are translated into the following ACL2 definitions. A few
minor \acl{define} options have been omitted for brevity. The \syn{connected} definition
is enabled based on an ad hoc simplicity heuristic.
\begin{bacl}
(define connected (e1 e2)
  :enabled t
  :returns (b booleanp :hyp :guard)
  (and (edge-p e1) (edge-p e2)
       (equal (edge->p2 e1)
              (edge->p1 e2)))
  ///
  (defret connected-ensures
    (and (booleanp b))
    :hyp :guard)
  (defthm connected-implies
    (implies (connected e1 e2)
             (and (edge-p e1) (edge-p e2)))))
(define path_p (edges)
  :measure (len edges)
  :returns (b booleanp :hyp :guard)
  (and (sequence[edge]-p edges)
       (or (< (len edges) 2)
           (and (connected (car edges)
                           (car (cdr edges)))
                (path_p (cdr edges)))))
  ///
  (defret path_p-ensures
    (and (booleanp b))
    :hyp :guard)
  (defthm path_p-implies
    (implies (path_p edges)
             (and (sequence[edge]-p edges)))))
 (defthm path_p_rest
   (implies (path_p edges)
            (path_p (cdr edges)))
   :hints (("Goal" :in-theory (enable path_p))))
\end{bacl}

The three main problem-definition functions:
\begin{lstlisting}
/* number of times edge0 crosses edges */
function crossings_count_aux(edge0: edge, edges: seq<edge>)
  assumes path_p(edges)
  returns (n: int) ensures n >= 0 {
  if (is_empty(edges)) {
    return 0;
  }
  else {
    if (edges_intersect(edge0, first(edges))) {
      return 1 + crossings_count_aux(edge0, rest(edges));
  }
  else {
    return crossings_count_aux(edge0, rest(edges));
  }}
}
function crossings_count(p: point, polygon: seq<point>)
  assumes simple_polygon(polygon)
  returns (n: int) ensures n >=0 {
  let pm:point = point(x=max_x(polygon) + 1, y=p.y);  /* pm is outside polygon */
  let e:edge = edge(p1 = p, p2 = pm);
  return crossings_count_aux(e,edges(polygon));
}
/* Top-level function */
function point_in_polygon(p: point, polygon: seq<point>)
  assumes simple_polygon(polygon)
  returns (b: bool) {
  return odd(crossings_count(p,polygon));
}
\end{lstlisting}

Most of the work is done by \syn{crossings_count_aux}, so we focus on that.
First, the type for \syn{seq<edge>} is defined in case it has not previously been
defined. A logical check if the guard is not satisfied is incorporated into the
base case of \syn{crossings_count_aux}.
\begin{bacl}
(fty::deflist sequence[edge]
  :elt-type edge)
(define crossings_count_aux ((edge0 edge-p)
                             (edges sequence[edge]-p))
  :measure (len edges)
  :guard (and (edge-p edge0)
              (sequence[edge]-p edges)
              (path_p edges))
  :returns (n natp :hyp :guard)
  (if (or (not (mbt (and (edge-p edge0)
                         (sequence[edge]-p edges)
                         (path_p edges))))
          (endp edges))
      0
    (if (edges_intersect edge0 (car edges))
        (+ 1 (crossings_count_aux edge0 (cdr edges)))
      (crossings_count_aux edge0 (cdr edges)))))
\end{bacl}

One common optimization is to convert the function to be tail recursive:
\begin{lstlisting}
function crossings_count_aux_1 =
  transform crossings_count_aux
    by tail_recursion {new_parameter_name = count}
\end{lstlisting}
This produces an ACL2 function that is back translated to Syntheto:
\begin{lstlisting}
function crossings_count_aux_1(edge0:edge,edges:seq<edge>,count:int)
  assumes path_p(edges)
  returns (n:int) {
  if (is_empty(edges)) {
     return count;
    }
    else {
      crossings_count_aux_1(edge0,rest(edges),
                              (edges_intersect(edge0, first(edges))
                                ? 1 : 0)
                              + count);
    }
}
\end{lstlisting}

The top-level function \syn{point_in_polygon} is given a polygon defined using a sequence
of vertices (points), from which a path is constructed to pass to
\syn{crossings_count}. We can exploit the existence of an isomorphism between sequences of
vertices and paths to avoid this extra step.
Using the isomorphism transformation requires
a proof of the isomorphism properties, which is done automatically.
\begin{lstlisting}
function crossings_count_aux_2 =
  transform crossings_count_aux_1
    by isomorphism {parameter = edges,
                      new_parameter_name = vertices,
                      old_type = path_p,
                      new_type = points2_p,
                      old_to_new = path_vertices,
                      new_to_old = path,
                      simplify = true}
\end{lstlisting}

This isomorphism transform produces the following ACL2 function:
\begin{bacl}
(defun crossings_count_aux_2 (edge0 vertices count)
  (declare (xargs :guard (and (points2_p vertices)
                              (edge-p edge0)
                              (sequence[edge]-p (path vertices))
                              (path_p (path vertices))
                              (natp count))
                  :measure (len (path vertices))))
  (and (mbt (points2_p vertices))
       (if (mbt (natp count))
           (if (or (not (mbt (edge-p edge0)))
                   (not (consp vertices))
                   (not (consp (cdr vertices))))
               count
             (crossings_count_aux_2
                edge0
                (rest vertices)
                (+ (if (edge_points_intersect (edge->p1 edge0) (edge->p2 edge0)
                                              (car vertices) (cadr vertices))
                       1 0)
                   count)))
          :undefined)))
\end{bacl}
This function is back-translated to Syntheto:
\begin{lstlisting}
function crossings_count_aux_2(edge0:edge,vertices:seq<point>,count:int)
  assumes (points2_p(vertices) && path_p(path(vertices)))
  returns (n:int) {
  if (is_empty(vertices) || is_empty(rest(vertices))) {
     return count;
   } else {
       crossings_count_aux_2(edge0,rest(vertices),
                               (edge_points_intersect
                                  (edge0.p1,edge0.p2,
                                   first(vertices),first(rest(vertices)))
                                 ? 1 : 0)
                               + count);
   }
}
\end{lstlisting}

\syn{point_in_polygon} calls \syn{odd} on the result of
\syn{crossing_count}. We can effectively push this call into the
\syn{crossings_count} computation by {\it wrapping} the function body of
\syn{crossings_count_aux_2} with the \syn{odd} function, then use finite
differencing to add a parameter that holds the current value of \syn{odd(count)}
and maintain its value by complementing its value on each call. Finally, the
value of \syn{count} is no longer used so it can be removed.
\begin{lstlisting}
function crossings_count_aux_3 =
  transform crossings_count_aux_2
    by wrap_output {wrap_function = odd}
function crossings_count_aux_4 =
  transform crossings_count_aux_3
    by finite_difference {expression = odd(count),
                             new_parameter_name = count_odd,
                             simplify = true}
function crossings_count_aux_5 =
  transform crossings_count_aux_4
    by drop_irrelevant_param {parameter = count}
\end{lstlisting}

Final optimized \syn{crossings_count_aux} function:
\begin{lstlisting}
function crossings_count_aux_5(edge0:edge,vertices:seq<point>,count_odd:bool)
  assumes (points2_p(vertices) && path_p(path(vertices)))
  returns (b:bool) {
  if (is_empty(vertices) || is_empty(rest(vertices))) {
     return count_odd;
   } else {
       crossings_count_aux_2(edge0,rest(vertices),
                               (edge_points_intersect
                                  (edge0.p1,edge0.p2,
                                   first(vertices),first(rest(vertices)))
                                 ? !count_odd : count_odd));
   }
}
\end{lstlisting}

The derivation is finalized by using the \syn{wrap_output} and \syn{simplify}
transformations to get the top-level functions to use the optimized
\syn{crossings_count_aux_5}. The previous transformations introduced rewrite rules that
replace the original functions by the transformed ones.
\begin{lstlisting}
function crossings_count_1 =
  transform crossings_count
    by wrap_output {wrap_function = odd,
                      simplify = true}
function point_in_polygon_final =
  transform point_in_polygon
    by simplify
\end{lstlisting}

The specification and transformations of this example are all performed in ACL2
and all proof obligations are discharged without any hints supplied by the
Syntheto user. There are 14 explicit theorems in the Syntheto specification in
order for these obligations to be proved and the transforms performed as shown.
Four of these were added to enable automatic proof of guard conditions. These are simple
and obvious enough that a relatively naive user could add them if we showed them the key
checkpoints of the failed proof back-translated to Syntheto, or it is possible the
system could automatically propose the theorems. Two of the explicit theorems are inversion
theorems necessary to prove the isomorphism in the transformation that produces
\syn{crossings_count_aux_2}. The final 7 explicit theorems are needed as rewrite rules to perform the
simplification steps in the derivations. These theorems were motivated by finding
expressions that could be optimized, in the results of the transformations.


\section{Related Work}
\label{sec:related}

We are not aware of other surface languages for ACL2
or for ACL2-based tools like APT.
Users of ACL2 and ACL2-based tools normally use the ACL2 language directly,
possibly taking advantage of macros to define and use custom notations
in the style of embedded domain-specific languages \cite{embedded-dsl}.
In contrast, Syntheto is a separate stand-alone language, with its own IDE,
and with ACL2 running behind the scenes.

We are also not aware of surface languages
for other theorem provers and related tools,
whose users normally use the provided language directly.
Some of these tools, e.g.\ Isabelle,
include IDEs that are close in style and functionality to the Syntheto IDE.

There exist other IDEs for ACL2. One IDE is the ACL2 Sedan \cite{acl2s}
running in Eclipse. This IDE communicates with ACL2 through stdin/stdout
rather than over the ACL2 bridge. There is also at least one implementation
of a Jupyter notebook interface for ACL2 \cite{gamboa-jupyter}.


The interrelated ideas of
program refinement,
program transformation,
and program synthesis
are not new
\cite{dijkstra-constructive,bmethod,zed,vdm,kids,specware-www},
and neither is their realization in a theorem prover like ACL2
\cite{popref,soft,apt-www,simplify,isodata}.
Syntheto builds on these ideas to provide
a more accessible language and a more feature-rich IDE.


\section{Future Work}
\label{sec:future}

Syntheto is currently just a prototype.
There are several directions in which it could be extended.

Type parameterization should be extended
from the built-in types for sets, sequences, etc.,
and the built-in operations on those types,
to user-defined types and operations.
 
More imperative-looking constructs could be added, such as loops,
provided that they can be translated to ACL2 and back with relative ease.
For example, loops could be represented as
tail-recursive functions of certain forms.

Since Syntheto currently covers
a relatively small selection of APT transformations,
another extension is to cover all the APT transformations.
The language also needs to be extended with a way to specify hints,
for both explicit and implicit theorems;
the challenge here is to provide a way to express them
using more a Syntheto vocabulary than an ACL2 vocabulary.

Since ACL2 has mainly been designed to be used by a human
interacting with a REPL, its output is unstructured text
sent to stdout.  It would be helpful to add more API-like
features to ACL2 for programmatically submitting prover
events and receiving structured data in reply.  That data
could be presented to the user of the IDE in a structured form.

Failed ACL2 proofs currently do not result in
much useful information sent back to the IDE.
We would like to experiment with
applying the back-translation from ACL2 to Syntheto
to the subgoal checkpoints that ACL2 displays when proof fails.
While this is not going to turn
the hard problem of general theorem proving into an easy one,
it is our experience that, at least in certain developments,
proofs sometimes fail for simple and fixable mistakes on the user's part,
and that the failed subgoal can often point the user to those mistakes,
if they are presented in an informative way.
This naturally leads to the idea of using counterexample generation,
which has a good precedent in the ACL2 Sedan.

The current IDE is very limited in capability, but VS Code
provides a rich customization environment.  Adding more functionality
to the IDE is a fruitful area to explore.
Examples are jumping from call site to definition
and having elidible inter-cell information returned by the prover.

Currently there is no facility for interactive execution of
Syntheto code.  It would be straightforward to add such a facility
to the front end, by running the code in the ACL2 server
as ACL2/Lisp code.

For delivering an application, we could add Syntheto constructs that translate to ACL2
calls of the Java or C code generators
\citeman{https://www.cs.utexas.edu/users/moore/acl2/manuals/latest/index.html?topic=JAVA____ATJ}{ATJ}
\citeman{https://www.cs.utexas.edu/users/moore/acl2/manuals/latest/index.html?topic=C____ATC}{ATC}.
This would enable the capability of synthesizing a
verified Java or C program that could be
distributed and run independently of ACL2.


\section*{Acknowledgements}

This work was partially sponsored by the Air Force Research Laboratory (AFRL)
and DARPA. The views, opinions, and/or findings expressed are those of the
author(s) and should not be interpreted as representing the official views or
policies of the Department of Defense or the U.S. Government.


\bibliographystyle{eptcs}
\bibliography{refs}


\end{document}